\documentclass[twocolumn]{aastex63}
\usepackage{acro}
\usepackage{amsmath}
\usepackage{multirow}
\usepackage{url}
\usepackage{color}

\newcommand\Tstrut{\rule{0pt}{2.9ex}}       % "top" strut
\newcommand\Bstrut{\rule[-1.3ex]{0pt}{0pt}} % "bottom" strut
\newcommand\TBstrut{\Tstrut\Bstrut}

\DeclareAcronym{AM}{
short = AM,
long = amplitude modulation,
class = abbrev
}

\DeclareAcronym{NS}{
short = NS,
long = neutron star,
class = abbrev
}

\DeclareAcronym{GW}{
short = GW,
long = gravitational wave,
class = abbrev
}

\DeclareAcronym{LMXB}{
short = LMXB,
long = low-mass X-ray binary,
class = abbrev
}

\DeclareAcronym{ScoX1}{
short = Sco~X-1,
long = Scorpius~X-1,
class = abbrev
}

\DeclareAcronym{CW}{
short = CW,
long = continuous wave,
class = abbrev
}

\DeclareAcronym{MDC}{
short = MDC,
long = mock-data challenge,
class = abbrev
}

\DeclareAcronym{SFT}{
short = SFT,
long = short Fourier transform,
class = abbrev
}

\DeclareAcronym{Tmax}{
short = $T_{\text{max}}$,
long = maximum time lag,
class = abbrev
}

\DeclareAcronym{FAP}{
short = FAP,
long = false alarm probability,
class = abbrev
}

\DeclareAcronym{SSB}{
short = SSB,
long = solar system barycenter,
class = abbrev
}

\DeclareAcronym{SNR}{
short = SNR,
long = signal to noise ratio,
class = abbrev
}

\DeclareAcronym{GR}{
short = GR,
long = general relativity,
class = abbrev
}

\DeclareAcronym{BH}{
short = BH,
long = black hole,
class = abbrev
}

\DeclareAcronym{CMB}{
short = CMB,
long = cosmic microwave background,
class = abbrev
}

\DeclareAcronym{PSD}{
short = PSD,
long = power spectral density,
class = abbrev
}

\DeclareAcronym{RMS}{
short = RMS,
long = root mean square,
class = abbrev
}

\DeclareAcronym{LIGO}{
short = LIGO,
long = Laser Interferometer Gravitational-Wave Observatory,
class = abbrev
}

\DeclareAcronym{FFT}{
short = FFT,
long = Fast Fourier Transform,
class = abbrev
}

\DeclareAcronym{LHO}{
short = LHO,
long = LIGO Hanford Observatory,
class = abbrev
}

\DeclareAcronym{CBC}{
short = CBC,
long = compact binary coalescence,
class = abbrev
}

\newcommand{\fref}[1]{Fig.~\ref{#1}}

\definecolor{Au}{rgb}{0.83, 0.69, 0.22}

\newcommand{\Tasc}{T_{\text{asc}}}
\newcommand{\Porb}{P_{\text{orb}}}

\newcommand{\tascAEIOtwo}{1178556229}
\newcommand{\TascOtwoAEISigma}{139}
\newcommand{\alfven}{Alfv\'{e}n}
\newcommand{\iotaorb}{\iota_{\textrm{orb}}}

\newcommand{\Tcoh}{{T_{\textrm{coh}}}}

\newcommand{\LivTsftStart}{1167984930}

\newcommand{\HanTsftSEnd}{1187733514}
\newcommand{\LivDurationhr}{2496}
\newcommand{\HanDurationhr}{2594}

\newcommand{\numofCluster}{32}
\newcommand{\numofKnownLine}{23}
\newcommand{\breakTBfreqO}{67.5}
\newcommand{\breakTBfreqE}{131.5}
\shorttitle{Search for continuous GWs from ScoX-1}
\shortauthors{Zhang et al.}

\begin{document}

\title{Search for Continuous Gravitational Waves from Scorpius X-1 in LIGO O2 Data}

\correspondingauthor{Yuanhao Zhang}
\email{yuanhao.zhang@aei.mpg.de}

\correspondingauthor{Maria Alessandra Papa}
\email{maria.alessandra.papa@aei.mpg.de}

\author{Yuanhao Zhang}
\affiliation{Max-Planck-Institut f\"{u}r Gravitationsphysik
  (Albert-Einstein-Institut), D-30167 Hannover, Germany}
\affiliation{Leibniz Universit\"at Hannover, 30167 Hannover, Germany}

\author{Maria Alessandra Papa}
%\email{maria.alessandra.papa@aei.mpg.de}
\affiliation{Max-Planck-Institut f\"{u}r Gravitationsphysik
  (Albert-Einstein-Institut), D-30167 Hannover, Germany}
\affiliation{Leibniz Universit\"at Hannover, 30167 Hannover, Germany}
\affiliation{Department of Physics, University of Wisconsin, Milwaukee, WI 53201, USA}

\author{Badri Krishnan}
\email{badri.krishnan@aei.mpg.de}
\affiliation{Max-Planck-Institut f\"{u}r Gravitationsphysik
  (Albert-Einstein-Institut), D-30167 Hannover, Germany}
\affiliation{Leibniz Universit\"at Hannover, 30167 Hannover, Germany}

\author{Anna L. Watts}
\email{A.L.Watts@uva.nl}
\affiliation{Anton Pannekoek Institute for Astronomy, University of Amsterdam, Postbus 94249, NL-1090 GE Amsterdam, the Netherlands}

%\collaboration{1}{(AAS Journals Data Scientists collaboration)}





\begin{abstract}
  We present the results of a search in LIGO O2 public data for continuous gravitational waves from the neutron star
  in the low-mass X-ray binary Scorpius X-1. We search for signals with $\approx$ constant frequency in the range 40-180 Hz. Thanks to the efficiency of our search pipeline we can use a long coherence time and achieve unprecedented sensitivity, significantly improving on existing results. This is the first search that has been able to probe gravitational wave amplitudes that could balance the accretion torque at the neutron star radius. Our search excludes emission at this level between \breakTBfreqO\, Hz and \breakTBfreqE\, Hz, for an inclination angle $44^\circ \pm 6^\circ$ derived from radio observations \citep{Fomalont:2001un}, and assuming that the spin axis is perpendicular to the orbital plane. If the torque arm is $\approx $ 26 km -- a conservative estimate of the \alfven\ radius -- our results are more constraining than the indirect limit across the band. This allows us to exclude certain mass-radius combinations and to place upper limits on the strength of the star's magnetic field. 
%  and in the central band beat that limit by a factor of about 2.  20\%
  We also correct a mistake that appears in the literature in the equation that gives the gravitational wave amplitude at the torque balance \citep{Abbott:2017hbu,Abbott:2019uwg} and we re-interpret the associated latest LIGO/Virgo results in light of this.  %\end{description}
\end{abstract}

\keywords{neutron stars --- gravitational waves --- continuous waves
  --- Sco X-1 --- accretion, accretion disks}  

\section{Introduction} \label{sec:intro}

Fast spinning neutron stars are promising sources of continuous gravitational waves in the frequency range 20 Hz - 2 kHz. The emission is typically generated by a non-axisymmetry in the star with respect to its rotation axis. The simplest example is the presence of an equatorial ellipticity that deforms the star into a triaxial ellipsoid rotating around the principal moment of inertia axis \citep{Jaranowski:1998qm}.

The strength of the gravitational wave signal is proportional to the ellipticity of the star. The {\it{maximum}} ellipticity that a neutron star could support before breaking has been estimated to lie in the $10^{-7}-10^{-5}$ range for neutron stars made of normal matter and a few orders of magnitude higher for exotic matter \citep{Horowitz09,JohnsonMcDaniel:2012wg,Baiko18,Gittins:2020cvx}.   The minimum ellipticity is harder to estimate:  we expect some ellipticity due to magnetic deformation, but the precise value depends strongly on the assumed magnetic field strength and configuration \citep[see for example][]{Haskell08,Mastrano11,Suvorov16}.  \citet{Woan18} have argued for a minimum ellipticity $\sim 10^{-9}$ based on the spin-down of millisecond pulsars (due to either magnetic field effects or some other source of ellipticity such as crustal deformation). 

For accreting neutron stars, the accretion process provides a potential additional source of asymmetry, particularly if accreting material is channeled unevenly onto the surface by the star's magnetic field.  This can lead to thermal and compositional gradients in the crust that generate a crustal `mountain' \citep{Bildsten:1998ey, 2000MNRAS.319..902U,Haskell06,Singh20}.  Accretion-induced deformation of the star's magnetic field might also result in asymmetries \citep{Melatos05,Vigelius09}. Accretion could also drive the excitation of some kind of internal oscillation that results in gravitational wave emission \citep{Andersson:1998qs, Haskell15review}.  Uncertainty about the accretion process and the stellar response makes it hard to compute firm estimates for the expected size of the resulting ellipticities, but they could be large enough for the resulting gravitational wave emission to be detectable with the current generation of detectors \citep{Lasky15review}.

What effect might such a gravitational wave torque have on an accreting neutron star? It has long been noted \citep{Papaloizou:1978,Wagoner:1984pv} 
that neutron stars in low mass X-ray binaries, in spite of having accreted matter for millions of years, spin well below the maximum possible spin frequency \citep{Cook94,Haensel09}, with the fastest accreting neutron star spinning at 620 Hz \citep{Hartman03,Patruno12,Watts12,Patruno:2017oum}.  Since gravitational wave torques scale with a high power of the frequency, as the spin rate increases, they naturally provide a mechanism that kicks-in more strongly than other mechanisms, preventing further spin-up.  This has led to the idea of torque balance, where gravitational wave and accretion torques reach equilibrium, preventing further spin-up and ensuring continuous gravitational wave emission \citep{Bildsten:1998ey}.   Indeed \cite{Gittins:2018cdw} have shown that a synthetic population of neutron stars evolved without the gravitational wave torque contribution does not produce the observed spin distribution. 

The accretion torque on a neutron star having mass $M$ is
\begin{equation}
\label{eq:accretionTorque}
N_{\rm{acc}} = \dot M \sqrt{GM r_m},
\end{equation} 
where $G$ is the gravitational constant, $r_m$ is the torque arm and $\dot{M}$ the accretion rate. The correct value to use for $r_m$ is not known a priori, but is typically assumed to be either the neutron star radius $R$ or the radius at which the star's magnetic field starts to disrupt the accretion flow.

The maximum accretion luminosity is ${{GM \dot M} \over R}$. If some fraction $X$ of this is radiated away by an X-ray flux $F_X$ observed at a distance $d$, then
\begin{equation}
\label{eq:Mdot}
X {{GM \dot M} \over R} = {4\pi d^2 F_X} \rightarrow \dot M={1\over X}{{4\pi d^2 F_X R}\over{GM}}.
\end{equation}

The gravitational wave intrinsic amplitude $h_0$  at a distance $d$, for a gravitational wave signal at twice the spin frequency of the star (which is the case if the ellipticity is caused by a magnetic or crustal mountain) and balancing the accretion torque, is 
\begin{equation}
\label{eq:h02Torque}
h_0^{torq.bal.} = \sqrt{{{5G}\over{2\pi^2 c^3}}{  \dot E_{\textrm{GW}}\over {d^2 f^2_{\textrm{GW}}}} }~~{\textrm{with}}~~\dot E_{\textrm{GW}}=\pi f_{\textrm{GW}} N_{\rm{acc}},
\end{equation}
where $f_{\textrm{GW}}$ is the gravitational wave frequency.
Substituting Eq.~\ref{eq:accretionTorque} and \ref{eq:Mdot} in Eq.~\ref{eq:h02Torque} one finds
\begin{equation}
 \begin{split}\label{eq:h0Torque}
&h_0^{torq.bal.} = \sqrt{ {{10}\over{Xc^3}}{ F_X R \over {f_{\textrm{GW}}}}  \sqrt{ {G r_m}\over M }}=\\
&= 3.4\times 10^{-26} 
\left(\frac{1.4\,M_{\odot}}{M}\right)^{\frac{1}{4}}
\left(\frac{r_m}{10\,\text{km}}\right)^{\frac{1}{4}} \times\\
 &\times   \left(\frac{F_X/X}{3.9\times
        10^{-7}\,\text{erg}~\text{cm}^{-2} \text{s}^{-1}}\right)^{\frac{1}{2}}
        \left(\frac{R}{10\,\text{km}}\right)^{\frac{1}{2}}
        \left(\frac{600\,\text{Hz}}{f_{\text{GW}}}\right)^{\frac{1}{2}}.        
        \end{split}
\end{equation}
We note that Eq.~15 in \cite{Abbott:2019uwg} and Eq.~10 in \cite{Abbott:2017hbu} are incorrect and yield the correct numerical value only if $r_m=R$. In those papers such mistake propagates to the \alfven\ radius torque balance amplitude curve of Fig.~5 (yellow curve in \citealt{Abbott:2019uwg}), which is over-estimated. This in turn makes it look like the constrained inclination angle upper limits from that search (for $\iota = \iotaorb\ \approx 44^\circ$) probe the \alfven\ radius torque balance limit, when in fact they do not.

For ease of notation we define 
  \begin{equation}
\label{eq:scaling}
   \begin{cases}
  M_1= \left( {M} \over {1.4\,M_{\odot}}\right)\\
  R_1=  \left( \frac{R}{10\,\text{km}} \right)\\
  r_{m1}=  \left( \frac{r_m}{10\,\text{km}} \right)\\
  B_1=  \left( {{B}\over{10^9 ~\textrm{G}}}\right)\\
  %f_{\textrm{GW}_1}=  \left(\frac{600\,\text{Hz}}   {f_{\text{GW}}}\right)\\
  f_{\textrm{GW}_1}=  \left(    \frac{f_{\text{GW}}}{600\,\text{Hz}}   \right)\\
  F_{X1} = \left(\frac{F_X}{3.9\times 10^{-7}~\text{erg}~\text{cm}^{-2}  \text{s}^{-1}}\right)\\
   d_1= \left( d\over {2.8 \text{kpc }} \right)\\
   %\dot{M}_1 = {{\dot{M}}\over{{3\times 10^{-8}\textrm{M}_\odot/\textrm{yr}}}}
  \end{cases}
\end{equation}
and re-write Eq.~\ref{eq:h0Torque} as
\begin{equation}
 \label{eq:h0TorqueShort}
 \begin{split}
 h_0^{torq.bal.} = & \,3.4\times 10^{-26} \,X^{-{1\over 2}} \\
& M_1^{-{1\over 4}} \,
{r_{m1}}^{\frac{1}{4}} \,
{{F_X}_1}^{\frac{1}{2}} \,
       {R_1}^{\frac{1}{2}} \,
       {{f_{\text{GW}}}_1}^{-{1\over 2}} .
       \end{split}
 \end{equation}

\ac{ScoX1} is the brightest persistent X-ray source after the Sun and hence, given the scaling of gravitational wave amplitude with X-ray flux,  it is a particularly promising continuous wave source. The flux value of  ${3.9\times 10^{-7}\,\text{erg}/\text{cm}^2/\text{s}}$ used in Eq.~\ref{eq:h0Torque} is the long-term average X-ray luminosity of \ac{ScoX1} measured from Earth \citep[see][for details of how this value was derived\footnote{The flux of \ac{ScoX1} during the O2 observations was comparable to the earlier observations used to generate the flux estimate, see \url{http://maxi.riken.jp/star_data/J1619-156/J1619-156.html}.}]{Watts:2008qw}.  This value yields a torque balance $h_0$ well within the reach of searches for continuous waves from known pulsars \citep{Authors:2019ztc,Abbott:2020lqk,Nieder:2019cyc,Nieder:2020yqy}.

Many searches have targeted continuous gravitational wave emission from \ac{ScoX1}
\citep[only since 2017]{Meadors2017,LVC_O1_Viterbi,LVC_O1_radiometer,Abbott:2017mwl,Abbott:2019uwg}, but none have yet been sensitive enough to probe the torque balance amplitudes of Eq.~\ref{eq:h0Torque}. This is because in contrast to the known pulsars targeted in \citet{Authors:2019ztc,Abbott:2020lqk,Nieder:2019cyc,Nieder:2020yqy}, the rotation frequency and frequency derivative of the \ac{ScoX1}-neutron star, as well as some binary parameters, are unknown. This means that a broad range of waveforms must be tested against the data, and this degrades the attainable sensitivity, through the increased trials factor. 

Another aspect that makes the  \ac{ScoX1} signal search challenging is its computational cost: as illustrated in \cite{Watts:2008qw} our ignorance of the system parameters results in a parameter space so broad that the most sensitive search method, a coherent matched filter over the entire observation time, is computationally prohibitive. This is a frequent predicament in searches for continuous gravitational waves and the standard solution is to adopt semi-coherent search methods, where one trades sensitivity in favour of computational efficiency \citep{ScoX1MDC,Dergachev:2019wqa}.

In semi-coherent searches the observation time is partitioned in segments spanning the same duration. If data from several instruments is used, these  partitions are $\approx$ coincident in time. The most important quantity is the duration of such partitions, $T_{\textrm{coh}}$. The larger $T_{\textrm{coh}}$ is, the more sensitive and the more computationally expensive the search is going to be.

We use for this search a cross-correlation method
\citep[and references therein]{LMXBCrossCorr}. Thanks to the much improved
computational efficiency of our new search \citep{CCResampMeadors}, we
are able to use a significantly longer $T_{\textrm{coh}}$ than ever
used before and reach unprecedented levels of sensitivity.  In
particular for the first time a search is sensitive to signals at the
torque balance limit at both the stellar radius and for reasonable estimates of the magnetospheric radius.

\section{The Search}
\label{sec:search}

We use LIGO O2 open data from the Hanford and Livingston detectors \citep{o2_data,Vallisneri:2014vxa} between GPS time \LivTsftStart\, (January 2016) and GPS time \HanTsftSEnd\, (August 2016). Overall we have 5090 hours of data, \LivDurationhr\, from Livingston and \HanDurationhr\, from Hanford. 

We search for a nearly monochromatic signal from the neutron star in \ac{ScoX1} --  below we qualify this assumption further. At the detector the signal appears frequency-modulated due to the relative motion between the star and the detector, and amplitude-modulated due to the sensitivity-response of the detectors, which depends on the line-of-sight direction and hence for a fixed source changes with time.
If all the source parameters were known, the gravitational waveform at the detector would also be known, and the search would be a perfectly matched filter,  like those carried out for known pulsars. This is not the case.

The \ac{LMXB} \ac{ScoX1} consists of a $1.4_{-0.5}^{+1.4}M_{\odot}$ neutron star and a
$0.7_{-0.3}^{+0.8}M_{\odot}$ companion star \cite[95\% confidence intervals]{ScoX1ParamRev}. No accretion-powered pulsations or thermonuclear burst oscillations have been so far detected from the neutron star, so its spin frequency is unknown. The orbital parameters projected semi-major axis, $a\sin{i}$, time of ascending nodes, $T_{\rm{asc}}$, 
%$T_{\rm{asc}}$(measured in the solar system barycenter) 
and orbital period, $P_{\rm{orb}}$, are constrained within ranges larger than our search resolution on those parameters, so these need to be explicitly searched \citep{ScoX1ParamRev}.

The search parameters are given in Table~\ref{tab:params}. We search for gravitational wave signal frequencies between 40 Hz and 180 Hz. The computational cost per unit frequency interval is smaller at lower frequencies, so concentrating computational resources in the lower frequency range makes for the highest return in sensitivity. In fact this is the frequency range in which we can match the torque balance limit, even with an unrestricted prior on the star's inclination angle. 

We do not explicitly search over frequency derivatives, reflecting the assumption that the system is close to equilibrium. With our search set-up we have measured an average loss in SNR at the 15\% level for gravitational wave first frequency derivative $|\dot f_{\textrm{GW}}| \simeq  2 \times 10^{-13}$ Hz/s. This sets the scale for the maximum rate of change of the spin frequency that would not affect our ability to detect a signal, at $|\dot f_{\textrm{spin}}| \lesssim$  $1 \times 10^{-13}$ Hz/s. We recall that for crustal mountains $f_{\textrm{GW}}=2 f_{\textrm{spin}}$.

The orbital parameter ranges are taken from Table 2 of 
\cite{ScoX1ParamRev}. $T_{\rm{asc}}$ is propagated to \tascAEIOtwo\, GPSs which is $\approx$ the weighted middle of the LIGO data observation span. We note that this is 206 epochs after the $T_{\rm{asc}}$ in \citep{Abbott:2019uwg}. Following Eq.~5 of \cite{Galloway2014}, we expand the uncertainty associated with $T_{\rm{asc}}$ to
\TascOtwoAEISigma\, seconds and then consider the 3$\sigma$ confidence interval.

The grid spacings in every dimension $d\lambda$ are chosen so that the loss due to signal-template mismatch is at the $m=25\%$ level. The spacings are estimated based on the metric $g_{\lambda\lambda}$ as  $d\lambda = \sqrt{\frac{m}{{g_{\lambda\lambda}}}}$. Expressions for the metric can be found in 
\cite{LMXBCrossCorr}. This approach results in an overestimate of the actual mismatch \citep{Allen:2019vcl}, and in fact we measure an overall average SNR loss of 16\%.
The grid spacings are given in Table~\ref{tab:params}. 

Our search employs a fixed $T_{\textrm{coh}}\simeq 19$ hrs, which is a factor of $4.5$ (10) longer than the longest (shortest) baseline used by \cite{O1KnownPulsar}.
This choice, enabled by the efficiency of our code (see Section \ref{sec:intro}) is the reason for the higher sensitivity of our search.

\begin{deluxetable}{c c c}
  \tablewidth{\textwidth}
  \tablecaption{Waveform parameter ranges
    \label{tab:params}} 
    \tablehead{ \colhead{Parameter} & \colhead{Range} & \colhead{Grid spacing} } 
    \startdata 
    \TBstrut $f_{\textrm{GW}}$ (Hz) & $[40, 180]$ & $\sim2\times 10^{-6}$ \\
  \TBstrut $a\sin i$ (lt-s) & $[1.45, 3.25]$ & $\sim\frac{0.17 ~ [{\textrm{lt-s}\,{\textrm{Hz}}}]}{f_{\textrm{GW}}}$ \\
  \TBstrut $\Tasc$ (GPS s)\tablenotemark{a} & $ \tascAEIOtwo\pm3\times\TascOtwoAEISigma$ & $\sim \frac{1576\, [{\textrm{lt-s}}]}{f_{\textrm{GW}} \,a\sin i}$ \\
  \TBstrut $\Porb$ (s) & $68023.86\pm 3\times 0.04$ & $\sim\frac{18\,[{\textrm{lt-s}}]}{f_{\textrm{GW}} \,a\sin i}$ \\
   & & \\
  \enddata 
%  \tablecomments{$\pm 3\sigma$ of the observational
%    uncertainties are covered for $\Tasc$ and $\Porb$.}
  %

  \tablenotetext{a}{Time of ascension has been propagated to May 11
    16:43:31 UTC 2017, close to the weighted-middle of the gravitational wave data, in order to make the metric approximately
    diagonal (\cite{LMXBCrossCorr}). The relation between $\Tasc$ and
    the epoch of inferior conjunction of the companion star $T_0$
    presented in \cite{ScoX1ParamRev} is 
    $\Tasc = T_0 - \Porb /4$ (\cite{ScoX1MDC})}. 
\end{deluxetable}

%\section{PostProcessing} \label{sec:post}

We consider search results with detection statistic values above the expected Gaussian-noise fluctuations. 
%The threshold is set so that, in Gaussian noise, we would not expect any result to exceed the threshold. 
Since the number of searched waveforms increases with frequency, noise fluctuations alone produce higher fluctuations at higher frequencies. For this reason our threshold for candidate consideration is not constant but rather increases with frequency.  

We find over 97 million 
%\zhangcmt{\numofCandLevZero} 
results above the threshold. As
often happens in this type of search, these results are not
uniformly distributed in frequency but tend to come in groups, with
the elements of each group having similar signal frequency, and due to
the same root cause. We cluster these together and examine each
cluster. We find \numofCluster\, such groups, which we will refer to as ``outlier
clusters". \numofKnownLine\, of them are
associated with known spectral contaminations \citep{Covas:2018oik}. The rest of the clusters are discarded based on cross-checking the
multi-detector detection statistics with the single-detector
statistics: When an outlier is due to a disturbance in one of the
detectors, the single-detector statistics will often be larger than
the multi-detector one. On the contrary a signal produces a higher
value when the data from both detectors is used. Most of these discarded clusters 
also present a range of signal frequencies with enhanced values of the detection statistic, that is much larger than it 
would be for a signal. The complete list of outlier clusters is given in  Table~\ref{tab:outliers}. 

%\mapcomment{map very first  preliminary pass done  until here}

\section{Results} 
\label{sec:result}

\begin{figure*}[tbp]
  \centering
  \vspace{0.5cm}
  \includegraphics[width=0.9\textwidth]{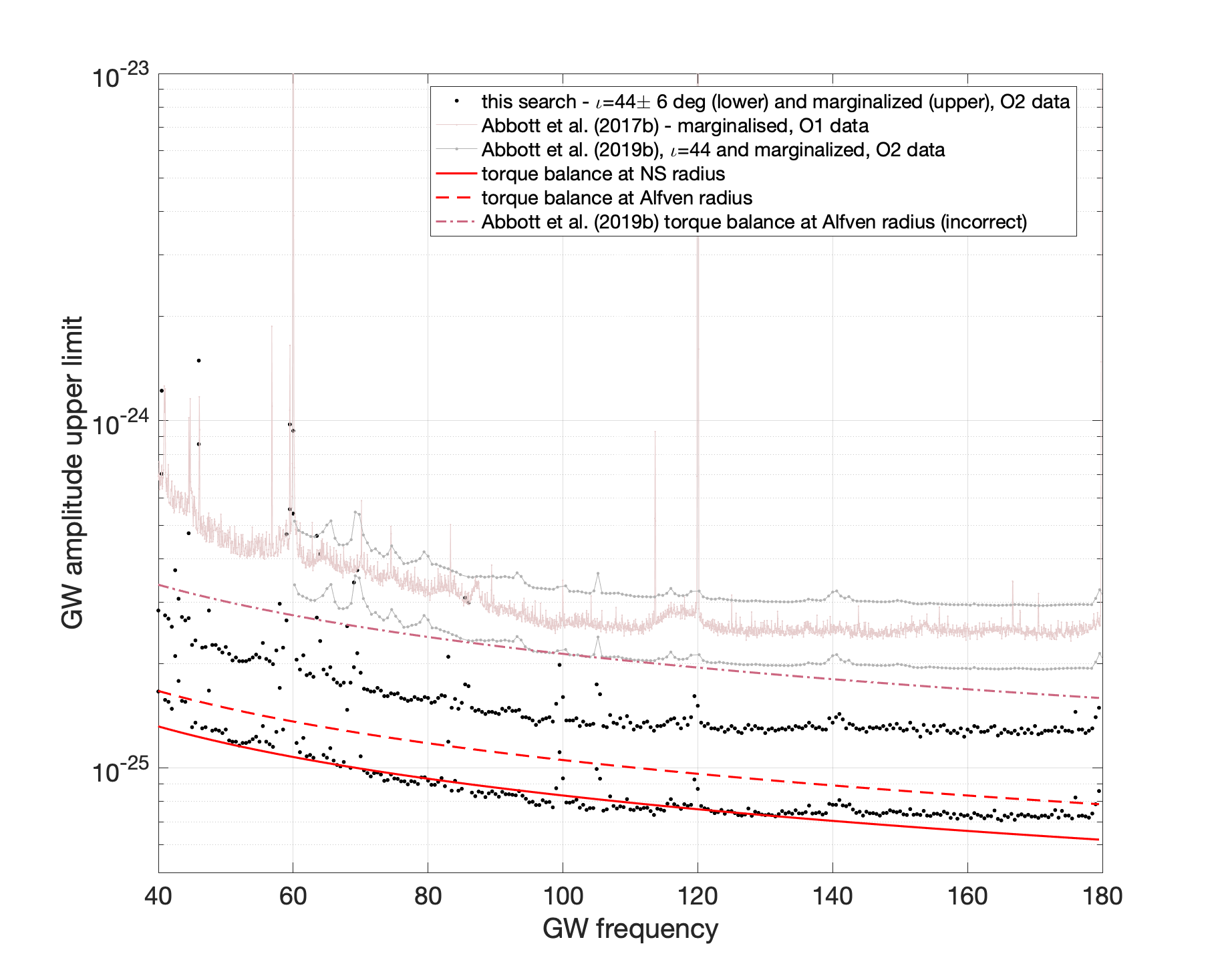}
  \caption{95\% confidence upper limits on the intrinsic gravitational wave amplitude in half-Hz bands. We assume orbital inclination at $44^\circ \pm 6^\circ$ (lower black points) and arbitrary inclination (upper black points). The lower dashed and solid curves are the torque-balance upper limits, based on estimates of the mass accretion rate and assuming the accretion torque to be at the neutron star radius (lower solid red curve) or at the  \alfven\ radius (upper dashed red curve). For comparison we show the upper limits from previous results (the three fainter upper curves) and draw the {\it incorrect} torque balance  \alfven\ radius upper limit that was reported (dash-dot line). }
  \label{fig:upperLimit}
\end{figure*}

\begin{figure}[tbp]
  \centering
  \vspace{0.5cm}
  \includegraphics[width=0.5\textwidth]{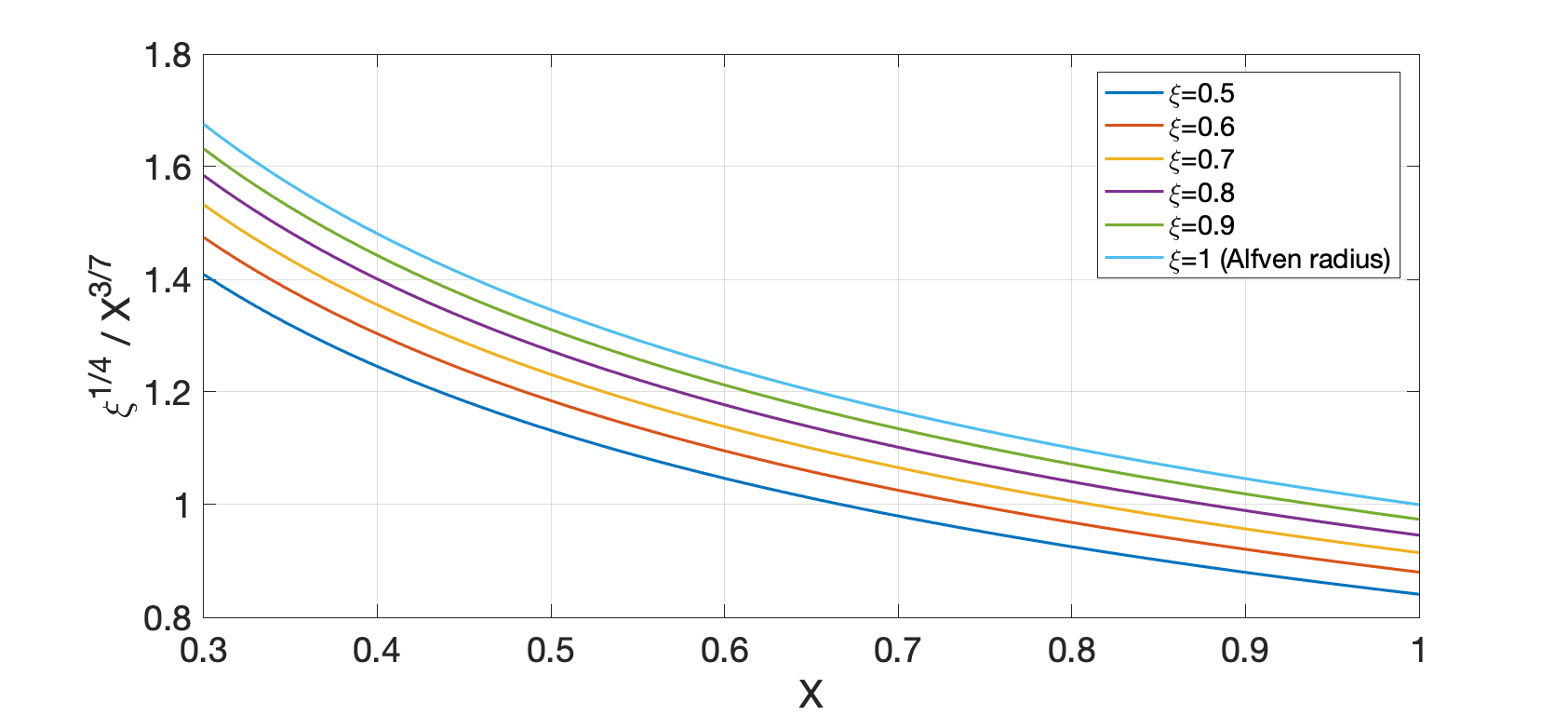}
  \caption{$h_0^{torq.bal.}(\xi, X)=h_0^{torq.bal.}(1,1){{\xi^{1/4}}\over X^{3/7}}$, and this multiplicative factor is shown here. }
  \label{fig:xiX}
\end{figure}

\subsection{Upper limits on GW amplitude}

As no significant candidate is found, we set upper limits at the  95\%  confidence level, on the gravitational wave intrinsic amplitude $h_0^{95\%}$ at the detectors, in half-Hz bands. 
%\fref{f:upperLimit}. 
The upper limits are determined by adding fake signals with a fixed amplitude $h_0$ to the data, and by measuring the detection efficiency, $C(h_0)$. The detection criterion is determined by the value of the detection statistic of the most significant result in the band. The procedure is repeated for various values of $h_0$ and a sigmoid fit is used to determine the value corresponding to 95\% confidence: $C(h_0^{95\%})=.95$ \citep{Fesik_2020}.

Two sets of upper limits are derived, reflecting two assumptions: 1) an arbitrary value of the inclination angle, with $\cos\iota$ uniformly distributed $-1 \leq \cos\iota \leq 1$, and 2) the inclination angle $\iota$ is equal to the orbital inclination angle and hence drawn from a Gaussian distribution with mean $44^\circ$ and standard deviation $6^{\circ}$ \citep{Fomalont:2001un,ScoX1ParamRev}. The latter scenario is equivalent to assuming that the spin axis of the neutron star is perpendicular to the orbital plane.  The $\iota=44^\circ\pm 6^\circ$ is a more favourable inclination than average for coupling to the gravitational wave detector \citep[see for instance Eq.s 21 and 22 ]{Jaranowski:1998qm} and the resulting upper limits are a factor $\approx$ 1.7 smaller than those for arbitrary orientation.

Both upper limits are plotted in \fref{fig:upperLimit} and provided in machine-readable format in \citet[and Suppl. Mat.]{O2CrossCorr-AEI}. For comparison \fref{fig:upperLimit} also shows upper limits from a previous cross-correlation search on O1 LIGO data \citep{Abbott:2017mwl} and from a recent Viterbi algorithm search on the same O2 data that we use \citep{Abbott:2019uwg}. The most sensitive of the two searches is the \cite{Abbott:2017mwl} search. It employed variable coherence lengths, with longer $\Tcoh$ in the low frequency range, which explains why at lower frequency it is comparatively more sensitive than at higher frequency.  The \citet{Abbott:2019uwg} search is less sensitive than a cross-correlation search but is more robust to deviations of the signal waveform from the assumed model \citep{Viterbi2016,ViterbiSideband}. In particular the method of \citet{Abbott:2019uwg} is robust with respect to loss of phase coherence in the signal.

One of the ways in which the signal could lose phase coherence with respect to the template waveforms of the search is through spin-wandering. This is a non-deterministic ``jitter" in the spin of star, caused, for instance, by small changes in the mass accretion rate. The resulting frequency variation depends on the accretion torque, hence on the spin frequency of the star, its moment of inertia, the ratio between the torque arm and the co-rotation radius and the mass accretion rate.  

Based on RXTE/ASM observations of Sco X-1, \cite{Mukherjee:2017qme} have explored different system-parameter combinations and the gravitational wave frequency changes that may accumulate over different observation periods, due to spin wandering. Their results indicate that, in our frequency range, the maximum frequency change during an observation time of $2\times 10^7$ s (our observation time) is less than $2\,\mu$Hz (our frequency resolution) for the vast majority of the simulated systems. This means that the sensitivity of this search should not be impacted by spin-wandering effects. 

Our results improve on existing ones by more than a factor $\approx$ 1.8. This is an extremely large sensitivity improvement in a large parameter space search like this one. For instance consider that in a broad all-sky search on O2 data, \cite{Pisarski:2019vxw} improve over the most sensitive results on O1 data \citep{Dergachev:2019wqa} by a factor of $\approx$ 1.1.

\subsection{Interpretation in terms of torque balance model}

Our results are also remarkable in absolute terms because they probe gravitational wave amplitudes that could support emission at the torque balance level. It is the first time that this milestone is reached. 

From Eq.s~\ref{eq:h0Torque} or \ref{eq:h0TorqueShort} we see that the torque balance gravitational wave amplitude depends on the torque arm and it is smallest at the star surface. If this minimum torque balance amplitude is {\it{larger}} than our $h_0^{95\%}$ upper limits it means that our search should have detected a signal; The fact that it has not, means that we can exclude such mass-radius combination:
\begin{equation}
\label{eq:RM}
 [ M_1^{-{1\over{4}}} R_1^{3\over4}]^{\textrm{excl}} \geq    {X^{1\over 2}} {{h_0^{95\%}}\over{3.4\times 10^{-26}}} f_{\textrm{GW}_1}^{\frac{1}{2}} F_1^{-{1\over 2}}.
\end{equation}
The lower panel of Figure  \ref{fig:Blimits} shows the mass-radius regions excluded by the $\iota \approx 44^\circ$ gravitational wave upper limits for $f_\text{GW}=117.5$ Hz and $X=1$.
% and $\xi=1$. 

If the torque arm is larger than the star radius, the torque balance amplitude increases, and our gravitational wave upper limits constrain the magnetic field strength of the mass-radius combinations not excluded by \ref{eq:RM}. We illustrate this point in the next paragraphs. 

We take the torque arm to be at the magnetospheric radius $r_m=\textrm{max}(\xi r_A, R)$, with $0.5\leq \xi \leq 1$ and $r_A$ the \alfven\ radius
\begin{equation}
\label{eq:alfven}
 %\begin{split}
   r_A = 
%   25.6 \,{B_1}^{4\over 7}\,
%   {R_1}^{12\over 7}\,
%   {{M}_1}^{-{1\over 7}} \,
%    {{\dot M}_1}^{-{2\over 7}} ~{\textrm{km}}   \\
%   %%
%   &= 
25.6 \,X^{2\over 7} \,{B_1}^{4\over 7}\,
   {R_1}^{10\over 7}\,
   {{M}_1}^{{1\over 7}} \,
    {{d}_1}^{-{4\over 7}} \,
    {{F_X}_1}^{-{2\over 7}} ~{\textrm{km}},
%!TEX encoding = UTF-8 Unicode \end{split}
\end{equation}
%The value used to scale $\dot{M}$ (as defined in Equation \ref{eq:scaling}), ${{3\times 10^{-8}\textrm{M}_\odot/\textrm{yr}}}$, corresponds to the measured ScoX-1 X-ray flux $F_X=3.9\times 10^{-7}~ \textrm{erg/cm}^2/\textrm{s}$ with $X=1$ , $R=10$ km, $M=1.4$ M$_\odot$ and $d=2.8$ kpc, in Eq.~\ref{eq:Mdot}. 
where $B_1$ is the normalised polar magnetic field strength, defined in Eq.~\ref{eq:scaling}. We note that in the gravitational wave literature the \alfven\ radius has often been placed at $35$ km, corresponding to $\dot M=10^{-8}\textrm{M}_\odot/\textrm{yr}$, or $X=0.3$ in Eq.~\ref{eq:Mdot}. The Eddington limit is at  $\dot M=2\times 10^{-8}\textrm{M}_\odot/\textrm{yr}$, for a fiducial 1.4 M$_\odot$ and 10 km radius neutron star. 

By combining Eq. \ref{eq:h0Torque} and Eq. \ref{eq:alfven} we find the torque-balance amplitude when $r_m \geq R$:
\begin{equation}
\label{eq:torqueBalanceRMB}
\begin{cases}
    h_0^{torq.bal.} =& \,4.3\times 10^{-26} 
  \,  {{\xi^{1\over 4} } \over {X^{3\over7} }}\\
    & \times {{F_X}_1}^{3\over 7}\,  {M_1}^{-{3\over{14}}} \,
    {R_1}^{6\over7} \, {B_1}^{1\over7} \,
      {d_1}^{-{1\over 7}}
    {f_{\textrm{GW}_1}}^{-{1\over 2}}\\
    B_1 \geq & 0.19 d_1 \xi^{-7/4} (F_1/X)^{1/2} (R_1^3 M_1)^{-1/4},\\
\end{cases}
\end{equation}
the last equation simply reflecting the condition $r_m \geq R$.  We note that $h_0^{torq.bal.}(\xi, X)=h_0^{torq.bal.}(1,1){{\xi^{1/4}}\over X^{3/7}}$ and this factor is plotted in \fref{fig:xiX} to aid evaluate how the torque balance amplitude changes under different assumptions for torque arm $r_m$ and the accretion luminosity.

%Again we stress that for the same mass-radius, this torque balance amplitude is never smaller than that with $r_m=R$. 

When this torque balance amplitude is {\it{larger}} than our $h_0^{95\%}$ upper limits it means that our search should have detected a signal; The fact that it has not, means that we can exclude the associated mass-radius-magnetic field strength combinations:
\begin{equation}
\label{eq:RMB}
  [M_1^{3\over{14}} R_1^{6\over7} B_1^{1\over7}]^{\textrm{excl}} \geq   { {X^{3\over 7}}\over {\xi^{1\over 4}}} {{h_0^{95\%}}\over{4.3\times 10^{-26}} }f_{\textrm{GW}_1}^{\frac{1}{2}} F_1^{-{3\over 7}} d_1^{1\over 7}.
 \end{equation}
This translates, for every mass-radius, into an upper limit on the magnetic field strength.

Figure \ref{fig:Blimits} shows the magnetic field upper limits from Eq.~\ref{eq:RMB} from the $\iota \approx 44^\circ$ gravitational wave upper limits for $f_\text{GW}=117.5$ Hz, $X=1$ and $\xi=1$, for different equations of state. The upper limits for different gravitational wave frequencies can be easily derived from the gravitational wave upper limit values using Eq.~\ref{eq:RMB}.  For the specific example shown in Figure \ref{fig:Blimits}, provided that the field is higher than $\sim 2\times 10^8$ G, the torque balance limit can be matched for all of the considered equations of state, but magnetic fields above $\sim 6\times 10^8$ G can be ruled out.

%The gravitational wave upper limits change with $f_{\textrm{GW}}$, which produces different $M-R-B$ constraints for different spins of the star. 
At \citet[and Suppl. Mat.]{O2CrossCorr-AEI} we provide plots like the one of Figure \ref{fig:Blimits} for gravitational wave frequencies in the searched range, at 2 Hz intervals. 

\begin{figure}[tbp]
  \centering
  \vspace{0.5cm}
  \includegraphics[width=0.49\textwidth]{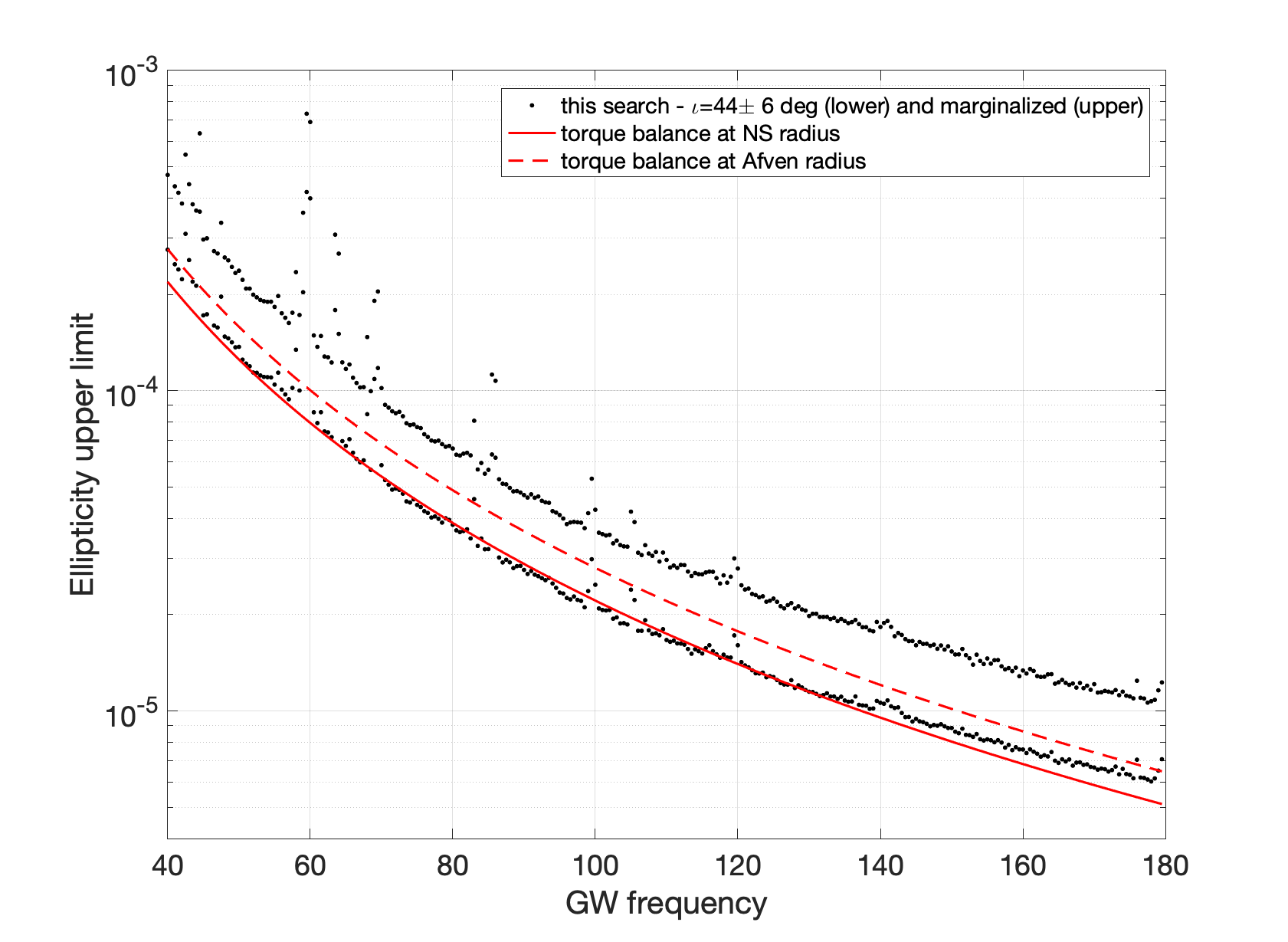}
  \caption{Upper limits on the ellipticity of the neutron star, derived from the gravitational wave intrinsic amplitude upper limits.}
  \label{fig:ellipticity}
\end{figure}
 
 The gravitational wave upper limits marginalised over all possible inclination angles lead to less stringent constraints on the physical parameters of the neutron star: the torque balance amplitude with torque arm at the neutron star surface is smaller than our upper limits for all equations of state, so no mass-radius combination can be ruled out. Torque-balance amplitudes larger than our upper limits can only be obtained for larger torque arms corresponding to magnetic field strengths $\gtrapprox 10^9$ G, which are higher than those expected from  \ac{LMXB}s.

If the gravitational wave signal is due to a triaxial ellipsoid rotating around a principal moment of inertia axis $I$, say along the $\hat{z}$ axis, the gravitational wave intrinsic amplitude $h_0$ is proportional to the ellipticity $\varepsilon$ of the star:
\begin{equation}
\label{eq:epsilon}
\begin{cases}
h_0 = \frac{4\pi^2 G}{c^4} {{I \varepsilon f^2_{\rm{GW}} }\over d}\\
\textrm{with} ~~ \varepsilon={{I_{xx}-I_{yy}}\over I}
\end{cases}
\end{equation}
We convert the $h_0^{95\%}$ upper limits into ellipticity upper
limits with Eq.~\ref{eq:epsilon}, with $d = 2.8 \text{kpc}$ and a fiducial value of $I = 10^{38} \text{kg} ~\text{m}^2$.  We also derive the ellipticity required for torque balance under the two previous assumptions on the lever arm. All these quantities are plotted in Fig.~\ref{fig:ellipticity}, as a function of the gravitational wave signal frequency.

Above $\sim$ 90 Hz the values of the ellipticity that we are exploring are a few $\times 10^{-5}$ and smaller. Deformations which are this large may be sustained by a neutron star crust \citep{JohnsonMcDaniel:2012wg}, although very recent work suggests that the maximum deformations may be smaller \citep{Gittins:2020cvx}.

\section{Discussion}
\label{sec:discussion}

%\com{Some thoughts on implications, which I think should be in the 'model-dependent' results sections:}
\begin{figure*}[tbp]
\centering
\vspace{0.5cm}
\includegraphics[width=0.9\textwidth]{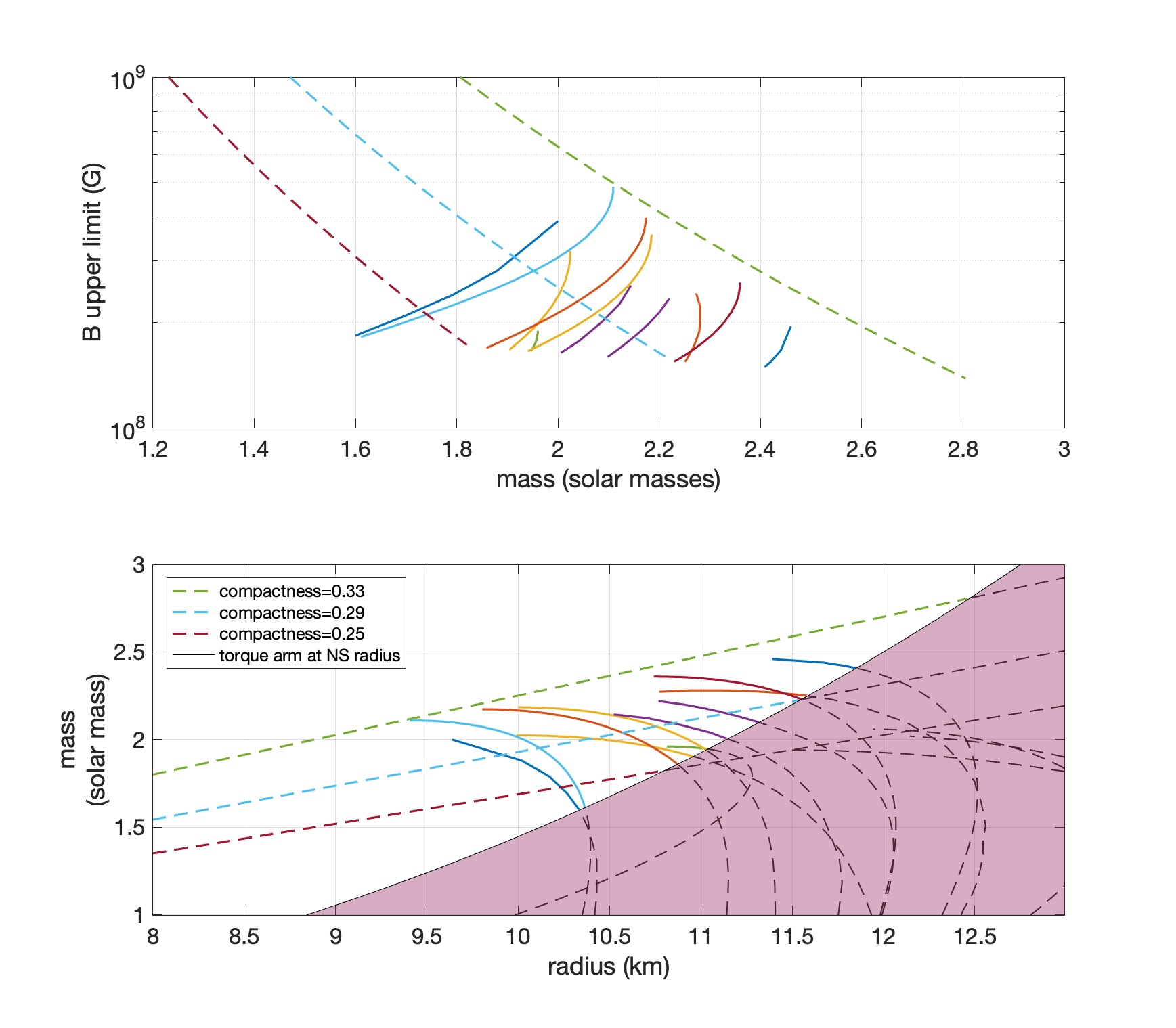}
\caption{We have assumed $f_{\text{GW}}=117.5$ Hz ($\approx 17$ ms spin period), $\iota=44^\circ\pm6^\circ$, $r_m=r_A ~(\xi=1)$, $X=1$ and torque balance. {\it{Top panel}}: the largest magnetic field consistent with our null result. The solid lines correspond to the equations of state from \citet[\texttt{http://xtreme.as.arizona.edu/NeutronStars/index.php/dense-matter-eos/}]{Ozel:2016oaf}. The dashed lines indicate stars of constant compactness $GM/Rc^2$ equal to 0.33 (upper), 0.29 (middle) and 0.25 (lower). We have considered masses in the range 1 to 3 M$_\odot$, radii between 8 and 13 km and we have dropped any equation of state with a maximum mass lower than 1.9 M$_\odot$, consistent with observations \citep{Antoniadis:2013pzd,Cromartie:2019kug} and with estimates from dense matter theory and experiment, \citep[see for example][]{Hebeler13,Kurkela14}. The lowest value of $B$ for each curve corresponds to $B(r_m=R)$. {\it{Lower panel}}: mass-radius relations for the equations of state considered above (solid lines) and for star configurations of constant compactness (dashed lines). The line that delimits the shaded region shows the mass-radius combinations that satisfy Eq.~\ref{eq:RM}, i.e. that are consistent with our upper limits when $r_m = R$. Below the shaded region the torque balance gravitational wave amplitude with $r_m = R$, is larger than our upper limits, so these configurations are excluded by our null results. Above the shaded region the torque balance gravitational wave amplitude with $r_m = R$, is smaller than our upper limits, so these configurations cannot be excluded if  $r_m = R$. If, however, $r_m > R$, i.e. a magnetic field above $\approx 2\times 10^8$ G, the corresponding torque balance becomes larger than our upper limits and this allows us to constrain the magnetic field (as shown in the top panel). }
\label{fig:Blimits}
\end{figure*}

This search has placed upper limits on stable GW emission that are tighter than the level predicted by torque balance models for \ac{ScoX1}, for $\iota\sim 44^\circ$.   This conclusion is robust to spin wandering at the level expected for this source. If the accretion torque is applied at the neutron star surface, the GW frequency range for which the torque balance limit is beaten is between \breakTBfreqO-\breakTBfreqE\, Hz, for a 1.4 M$_\odot$ and 10 km radius fiducial star.   If on the other hand the torque is applied at a magnetospheric radius at 25.6 km (see Eq.s~\ref{eq:scaling} and \ref{eq:alfven}), then the range for which the limit is beaten is the entire searched range, 40-180 Hz, for the fiducial star, as shown in \fref{fig:upperLimit}.

If we consider a wider range of masses and radii, consistent with our current best understanding of viable equation of state models, we are able to place constraints on mass-radius-magnetic field strength combinations:
\begin{itemize}
\item independently of the magnetic field value, our results exclude certain mass-radius combinations. Our tightest limits come for spin periods of $\sim $ 21 ms ($f_{\textrm{GW}}\sim 96$ Hz, at twice the spin frequency) with a narrow range of allowed masses extending only between 1.9-2.2 M$_\odot$ and magnetic fields larger than $\sim 3\times 10^8$ G being ruled out for all considered equations of state.
\item if the magnetic field is larger than $0.19 d_1 \xi^{-7/4} (F_1/X)^{1/2} (R_1^3 M_1)^{-1/4}$ ($r_m > R$) we can place upper limits on the magnetic field strength. The upper limit on the magnetic field is highest for the highest mass in the range. Stiffer equations of state have a smaller range of masses (and magnetic field strengths) for which balance can still be possible at the level of our upper limits, than softer equations of state. We find that the field must be smaller than $(4-10) \times 10^8$ G, depending on frequency (but excluding too disturbed frequency ranges, e.g. 60 Hz), for all equations of state models considered.
\end{itemize}

It is the first time that constraints on the magnetic field, mass and radius are obtained through continuous wave observations. This is interesting because the magnetic field is in general very poorly constrained and because observations like these probe mass-radius and magnetic fields through an entirely different mechanism than gravitational wave binary inspiral signals \citep[see e.g.][]{PhysRevX.9.011001,Capano:2019eae}. 

If the spin of \ac{ScoX1} is such that it is in the range where the limit is beaten 
(half the GW frequency for mountain models) and torque balance applies, this means that GW emission is not strong enough to balance the assumed accretion torque. This implies that the accretion torque must be less strong than predicted by the models presented in this paper, which could happen if, for example, strong radiation pressure modifies the structure of the inner disk \citep{Andersson05} or due to the effect of winds \citep{Parfrey16}.  

The result also puts limits on the size of thermal/compositional crustal or magnetic mountains in \ac{ScoX1}.  Limits can also be placed on internal oscillation amplitude for models where that is the mechanism that provides the GW torque (for a different range of spin frequencies since the relationship between spin and GW frequency is different for mode models).   

An alternative, of course, is that the spin of \ac{ScoX1} is outside the range searched, perhaps at higher frequencies more in line with the spin rates measured for the accretion-powered millisecond pulsars and thermonuclear burst oscillation sources \citep{Patruno12,Watts12}.  

At 1 kHz the torque balance upper limit for the ``fiducial star" is between 2.6-3.3 $\times 10^{-26}$, for $r_m\in [10-25.6]$ km.  This is about a factor of 5-10 lower than a signal that we could detect at that frequency with a search like this one -- the variation depending on the torque arm and on the inclination angle, and assuming that one could actually perform this search at such high frequencies. %Bridging this gap in sensitivity is going to be very hard: 
If all the parameters of \ac{ScoX1} were known, a search on the same O2 data as used here could probably detect signals at 2.6-3.3 $\times 10^{-26}$. 
We are however quite far from having a complete ephemeris for  \ac{ScoX1}. The next best thing would be to know the rotation frequency of the neutron star. The reason is that the torque balance amplitude decreases with frequency (so the sensitivity requirement increases, to match the torque balance limit), and the sensitivity of the searches decreases with frequency due to the shot noise in the detectors and to the increased template resolution per-Hz searched. These factors make it difficult to search very broad frequency bands. If it were possible to identify the spin frequency, for example via the detection of weak or intermittent pulsations (a major goal for future large-area X-ray telescopes \citealt{extp,strobex}), we might be able to carry out a search like this one, that could begin to probe the torque balance limit when the noise level at 1 kHz reaches its design value of $\sim 5.5\times 10^{-24}\, 1/\sqrt{\textrm{Hz}}$ \citep{Aasi:2013wya} and with $\sim$ two years of data.

\section{ACKNOWLEDGMENTS}

The computation of the work was run on the ATLAS computing cluster at
AEI Hannover \cite{Atlas} funded by the Max Planck Society and the
State of Niedersachsen, Germany.  A.L.W. acknowledges support from ERC Consolidator Grant No.~865768 AEONS (PI: Watts). \\
This research has made use of data, software and/or web tools obtained from the LIGO Open Science Center (\url{https://losc.ligo.org}), a service of LIGO Laboratory, the LIGO Scientific Collaboration and the Virgo Collaboration.  LIGO is funded by the U.S. National Science Foundation. Virgo is funded by the French Centre National de Recherche Scientifique (CNRS), the Italian Istituto Nazionale della Fisica Nucleare (INFN) and the Dutch Nikhef, with contributions by Polish and Hungarian institutes.

\appendix
\section{Outlier table}

\begin{deluxetable*}{cccc}
  \tablewidth{\textwidth}
  \tablecaption{Table of outlier clusters, as described in the text.
    \label{tab:outliers}} 
 \tablehead{ \colhead{Cluster ID} & \colhead{freq [Hz] } & \colhead{detection statistic} & \colhead{description} } 
    \startdata 
      0  & 40.883985  & 180.55      & known line in H1/L1 \& too broad in freq.                              \\%40.87, 40.88, 40.90, 40.92
      1  & 42.852488  & 19.05       & known line in H1/L1 \& too broad in freq.                              \\%42.85, 42.86, 42.87
      2  & 43.338994  & 12.99       & fails single/multi-detector statistic comparison \& too broad in freq. \\
      3  & 44.524761  & 35.08       & known line in H1/L1 \& too broad in freq.                              \\%44.50, 44.55
      4  & 46.089393  & 415.72      & fails single/multi-detector statistic comparison \& too broad in freq. \\
      5  & 47.679448  & 15.54       & known line in H1/L1 \& too broad in freq.                              \\%47.6833
      6  & 55.565183  & 11.96       & known line in H1/L1                                                      \\%55.5559, 55.5562
      7  & 58.188448  & 17.06       & fails single/multi-detector statistic comparison \& too broad in freq. \\%58.2854
      8  & 59.516247  & 42.19       & known line in H1/L1 \& too broad in freq.                              \\%59.50
      9  & 60.005969  & 185.90      & known line in H1/L1 \& too broad in freq.                              \\%60 Hz power line
     10  & 61.814927  & 11.50       & known line in H1/L1 \& too broad in freq.                              \\%61.8009, 61.8012
     11  & 63.994606  & 68.80       & known line in H1/L1 \& too broad in freq.                              \\%63.9987, 64.0000
     12  & 64.025584  & 13.24       & same as cluster 11 \& too broad in freq.                               \\
     13  & 64.279235  & 14.38       & fails single/multi-detector statistic comparison \& too broad in freq.  \\
     14  & 68.463375  & 21.05       & known line in H1/L1 \& too broad in freq.                              \\%68.50
     15  & 68.492385  & 18.08       & known line in H1/L1 \& too broad in freq.                              \\
     16  & 69.564208  & 41.68       & known line in H1/L1 \& too broad in freq.                              \\%69.50
     17  & 69.649598  & 13.55       & known line in H1/L1 \& too broad in freq.                              \\%69.6044
     18  & 69.764916  & 16.83       & known line in H1/L1 \& too broad in freq.                              \\%69.7752, 69.7756
     19  & 83.301663  & 18.59       & fails single/multi-detector statistic comparison \& too broad in freq.  \\
     20  & 85.989651  & 43.63       & known line in H1/L1 \& too broad in freq.                              \\%85.9987, 86.0000
     21  & 99.332771  & 11.80     & fails single/multi-detector statistic comparison                         \\%99.30 - 99.40 Hz      \\
     22  & 99.382711  & 11.54   & fails single/multi-detector statistic comparison                         \\
     23  & 99.984856  & 18.68       & known line in H1/L1 \& too broad in freq.                              \\%99.9760, 99.9987, 100.0000, 100.0011
     24  & 105.249255 & 15.40       & fails single/multi-detector statistic comparison \& too broad in freq.  \\%105.15 - 105.40 Hz    \\
     25  & 105.467399 & 12.97       & known line in H1/L1 \& too broad in freq.                              \\%105.50
    %26  & 105.519829 & 11.49       & disturbed band                                                           \\
     26  & 105.603586 & 13.66       & known line in H1/L1 \& too broad in freq.                              \\%105.6596, 105.6602, 105.7495, 105.7493
     27  & 119.886974 & 13.63       & known line in H1/L1 \& too broad in freq.                              \\%120 Hz power line harmonic              \\
     28  & 119.934821 & 13.04       & known line in H1/L1 \& too broad in freq.                              \\
     29  & 120.002786 & 11.84       & known line in H1/L1 \& too broad in freq.                              \\
     30  & 176.308216 & 13.28       & fails single/multi-detector statistic comparison \& too broad in freq. \\
     31  & 179.994556 & 12.76       & known line in H1/L1
 \enddata 
\end{deluxetable*}

\bibliography{biblio}{}

\begin{thebibliography}{}
\expandafter\ifx\csname natexlab\endcsname\relax\def\natexlab#1{#1}\fi
\providecommand{\url}[1]{\href{#1}{#1}}
\providecommand{\dodoi}[1]{doi:~\href{http://doi.org/#1}{\nolinkurl{#1}}}
\providecommand{\doeprint}[1]{\href{http://ascl.net/#1}{\nolinkurl{http://ascl.net/#1}}}
\providecommand{\doarXiv}[1]{\href{https://arxiv.org/abs/#1}{\nolinkurl{https://arxiv.org/abs/#1}}}

\bibitem[{Abbott {et~al.}(2017{\natexlab{a}})}]{LVC_O1_Viterbi}
Abbott, B., {et~al.} 2017{\natexlab{a}}, Phys. Rev. D, 95, 122003,
  \dodoi{10.1103/PhysRevD.95.122003}

\bibitem[{Abbott {et~al.}(2018)}]{Aasi:2013wya}
---. 2018, Living Rev. Rel., 21, 3, \dodoi{10.1007/s41114-018-0012-9}

\bibitem[{Abbott {et~al.}(2019{\natexlab{a}})Abbott, Abbott, Abbott, Abraham,
  Acernese, Ackley, Adams, Adhikari, Adya, Affeldt, \& et~al.}]{Abbott:2019uwg}
Abbott, B., Abbott, R., Abbott, T., {et~al.} 2019{\natexlab{a}}, Physical
  Review D, 100, \dodoi{10.1103/physrevd.100.122002}

\bibitem[{Abbott {et~al.}(2019{\natexlab{b}})}]{Authors:2019ztc}
Abbott, B., {et~al.} 2019{\natexlab{b}}, Astrophys. J., 879, 10,
  \dodoi{10.3847/1538-4357/ab20cb}

\bibitem[{Abbott {et~al.}(2019{\natexlab{c}})}]{Pisarski:2019vxw}
---. 2019{\natexlab{c}}, Phys. Rev. D, 100, 024004,
  \dodoi{10.1103/PhysRevD.100.024004}

\bibitem[{Abbott {et~al.}(2017{\natexlab{b}})}]{Abbott:2017hbu}
Abbott, B.~P., {et~al.} 2017{\natexlab{b}}, Phys. Rev., D95, 122003,
  \dodoi{10.1103/PhysRevD.95.122003}

\bibitem[{Abbott {et~al.}(2017{\natexlab{c}})}]{LVC_O1_radiometer}
---. 2017{\natexlab{c}}, Phys.\ Rev.\ Lett., 118, 121102,
  \dodoi{10.1103/PhysRevLett.118.121102}

\bibitem[{Abbott {et~al.}(2017{\natexlab{d}})}]{Abbott:2017mwl}
---. 2017{\natexlab{d}}, Astrophys. J., 847, 47,
  \dodoi{10.3847/1538-4357/aa86f0}

\bibitem[{Abbott {et~al.}(2017{\natexlab{e}})Abbott, Abbott, Abbott, Abernathy,
  Acernese, Ackley, Adams, Adams, Addesso, Adhikari, \& et~al.}]{O1KnownPulsar}
Abbott, B.~P., Abbott, R., Abbott, T.~D., {et~al.} 2017{\natexlab{e}}, The
  Astrophysical Journal, 839, 12, \dodoi{10.3847/1538-4357/aa677f}

\bibitem[{Abbott {et~al.}(2019{\natexlab{d}})}]{PhysRevX.9.011001}
Abbott, B.~P., {et~al.} 2019{\natexlab{d}}, Phys. Rev. X, 9, 011001,
  \dodoi{10.1103/PhysRevX.9.011001}

\bibitem[{Abbott {et~al.}(2020)}]{Abbott:2020lqk}
Abbott, R., {et~al.} 2020.
\newblock \doarXiv{2007.14251}

\bibitem[{AEI(2017)}]{Atlas}
AEI. 2017, {The Atlas Computing Cluster},
  https://www.aei.mpg.de/24838/02\_Computing\_and\_ATLAS.
\newblock \url{https://www.aei.mpg.de/24838/02\_Computing\_and\_ATLAS}

\bibitem[{Allen(2019)}]{Allen:2019vcl}
Allen, B. 2019, Phys. Rev. D, 100, 124004, \dodoi{10.1103/PhysRevD.100.124004}

\bibitem[{{Andersson} {et~al.}(2005){Andersson}, {Glampedakis}, {Haskell}, \&
  {Watts}}]{Andersson05}
{Andersson}, N., {Glampedakis}, K., {Haskell}, B., \& {Watts}, A.~L. 2005,
  \mnras, 361, 1153, \dodoi{10.1111/j.1365-2966.2005.09167.x}

\bibitem[{Andersson {et~al.}(1999)Andersson, Kokkotas, \&
  Stergioulas}]{Andersson:1998qs}
Andersson, N., Kokkotas, K.~D., \& Stergioulas, N. 1999, Astrophys.\ J., 516,
  307, \dodoi{10.1086/307082}

\bibitem[{Antoniadis {et~al.}(2013)}]{Antoniadis:2013pzd}
Antoniadis, J., {et~al.} 2013, Science, 340, 6131,
  \dodoi{10.1126/science.1233232}

\bibitem[{{Baiko} \& {Chugunov}(2018)}]{Baiko18}
{Baiko}, D.~A., \& {Chugunov}, A.~I. 2018, \mnras, 480, 5511,
  \dodoi{10.1093/mnras/sty2259}

\bibitem[{Bildsten(1998)}]{Bildsten:1998ey}
Bildsten, L. 1998, Astrophys.\ J.\ Lett., 501, L89, \dodoi{10.1086/311440}

\bibitem[{Capano {et~al.}(2020)Capano, Tews, Brown, Margalit, De, Kumar, Brown,
  Krishnan, \& Reddy}]{Capano:2019eae}
Capano, C.~D., Tews, I., Brown, S.~M., {et~al.} 2020, Nature Astron., 4, 625,
  \dodoi{10.1038/s41550-020-1014-6}

\bibitem[{{Cook} {et~al.}(1994){Cook}, {Shapiro}, \& {Teukolsky}}]{Cook94}
{Cook}, G.~B., {Shapiro}, S.~L., \& {Teukolsky}, S.~A. 1994, \apjl, 423, L117,
  \dodoi{10.1086/187250}

\bibitem[{Covas {et~al.}(2018)}]{Covas:2018oik}
Covas, P., {et~al.} 2018, Phys. Rev. D, 97, 082002,
  \dodoi{10.1103/PhysRevD.97.082002}

\bibitem[{Cromartie {et~al.}(2019)}]{Cromartie:2019kug}
Cromartie, H., {et~al.} 2019, Nature Astron., 4, 72,
  \dodoi{10.1038/s41550-019-0880-2}

\bibitem[{Dergachev \& Papa(2019)}]{Dergachev:2019wqa}
Dergachev, V., \& Papa, M.~A. 2019, Phys. Rev. Lett., 123, 101101,
  \dodoi{10.1103/PhysRevLett.123.101101}

\bibitem[{Fesik \& Papa(2020)}]{Fesik_2020}
Fesik, L., \& Papa, M.~A. 2020, The Astrophysical Journal, 895, 11,
  \dodoi{10.3847/1538-4357/ab8193}

\bibitem[{Fomalont {et~al.}(2001)Fomalont, Geldzahler, \&
  Bradshaw}]{Fomalont:2001un}
Fomalont, E., Geldzahler, B., \& Bradshaw, C. 2001, Astrophys. J., 558, 283,
  \dodoi{10.1086/322479}

\bibitem[{{Galloway} {et~al.}(2014){Galloway}, {Premachandra}, {Steeghs},
  {Marsh}, {Casares}, \& {Cornelisse}}]{Galloway2014}
{Galloway}, D.~K., {Premachandra}, S., {Steeghs}, D., {et~al.} 2014,
  Astrophys.\ J., 781, 14, \dodoi{10.1088/0004-637X/781/1/14}

\bibitem[{Gittins \& Andersson(2019)}]{Gittins:2018cdw}
Gittins, F., \& Andersson, N. 2019, Mon. Not. Roy. Astron. Soc., 488, 99,
  \dodoi{10.1093/mnras/stz1719}

\bibitem[{Gittins {et~al.}(2020)Gittins, Andersson, \& Jones}]{Gittins:2020cvx}
Gittins, F., Andersson, N., \& Jones, D. 2020.
\newblock \doarXiv{2009.12794}

\bibitem[{{Haensel} {et~al.}(2009){Haensel}, {Zdunik}, {Bejger}, \&
  {Lattimer}}]{Haensel09}
{Haensel}, P., {Zdunik}, J.~L., {Bejger}, M., \& {Lattimer}, J.~M. 2009, \aap,
  502, 605, \dodoi{10.1051/0004-6361/200811605}

\bibitem[{{Hartman} {et~al.}(2003){Hartman}, {Chakrabarty}, {Galloway}, {Muno},
  {Savov}, {Mendez}, {van Straaten}, \& {Di Salvo}}]{Hartman03}
{Hartman}, J.~M., {Chakrabarty}, D., {Galloway}, D.~K., {et~al.} 2003, in
  AAS/High Energy Astrophysics Division \#7, AAS/High Energy Astrophysics
  Division, 17.38

\bibitem[{{Haskell}(2015)}]{Haskell15review}
{Haskell}, B. 2015, International Journal of Modern Physics E, 24, 1541007,
  \dodoi{10.1142/S0218301315410074}

\bibitem[{{Haskell} {et~al.}(2006){Haskell}, {Jones}, \&
  {Andersson}}]{Haskell06}
{Haskell}, B., {Jones}, D.~I., \& {Andersson}, N. 2006, \mnras, 373, 1423,
  \dodoi{10.1111/j.1365-2966.2006.10998.x}

\bibitem[{{Haskell} {et~al.}(2008){Haskell}, {Samuelsson}, {Glampedakis}, \&
  {Andersson}}]{Haskell08}
{Haskell}, B., {Samuelsson}, L., {Glampedakis}, K., \& {Andersson}, N. 2008,
  \mnras, 385, 531, \dodoi{10.1111/j.1365-2966.2008.12861.x}

\bibitem[{{Hebeler} {et~al.}(2013){Hebeler}, {Lattimer}, {Pethick}, \&
  {Schwenk}}]{Hebeler13}
{Hebeler}, K., {Lattimer}, J.~M., {Pethick}, C.~J., \& {Schwenk}, A. 2013,
  \apj, 773, 11, \dodoi{10.1088/0004-637X/773/1/11}

\bibitem[{{Horowitz} \& {Kadau}(2009)}]{Horowitz09}
{Horowitz}, C.~J., \& {Kadau}, K. 2009, \prl, 102, 191102,
  \dodoi{10.1103/PhysRevLett.102.191102}

\bibitem[{Jaranowski {et~al.}(1998)Jaranowski, Krolak, \&
  Schutz}]{Jaranowski:1998qm}
Jaranowski, P., Krolak, A., \& Schutz, B.~F. 1998, Phys.\ Rev.\ D., 58, 063001,
  \dodoi{10.1103/PhysRevD.58.063001}

\bibitem[{Johnson-McDaniel \& Owen(2013)}]{JohnsonMcDaniel:2012wg}
Johnson-McDaniel, N.~K., \& Owen, B.~J. 2013, Phys. Rev. D, 88, 044004,
  \dodoi{10.1103/PhysRevD.88.044004}

\bibitem[{{Kurkela} {et~al.}(2014){Kurkela}, {Fraga}, {Schaffner-Bielich}, \&
  {Vuorinen}}]{Kurkela14}
{Kurkela}, A., {Fraga}, E.~S., {Schaffner-Bielich}, J., \& {Vuorinen}, A. 2014,
  \apj, 789, 127, \dodoi{10.1088/0004-637X/789/2/127}

\bibitem[{{Lasky}(2015)}]{Lasky15review}
{Lasky}, P.~D. 2015, \pasa, 32, e034, \dodoi{10.1017/pasa.2015.35}

\bibitem[{LIGO(2019)}]{o2_data}
LIGO. 2019, The O2 Data Release, \url{https://www.gw-openscience.org/O2/},
  \dodoi{10.7935/CA75-FM95}

\bibitem[{{Mastrano} {et~al.}(2011){Mastrano}, {Melatos}, {Reisenegger}, \&
  {Akg{\"u}n}}]{Mastrano11}
{Mastrano}, A., {Melatos}, A., {Reisenegger}, A., \& {Akg{\"u}n}, T. 2011,
  \mnras, 417, 2288, \dodoi{10.1111/j.1365-2966.2011.19410.x}

\bibitem[{Meadors {et~al.}(2017)Meadors, Goetz, Riles, Creighton, \&
  Robinet}]{Meadors2017}
Meadors, G.~D., Goetz, E., Riles, K., Creighton, T., \& Robinet, F. 2017,
  Phys.\ Rev.\ D., 95, 042005, \dodoi{10.1103/PhysRevD.95.042005}

\bibitem[{Meadors {et~al.}(2018)Meadors, Krishnan, Papa, Whelan, \&
  Zhang}]{CCResampMeadors}
Meadors, G.~D., Krishnan, B., Papa, M.~A., Whelan, J.~T., \& Zhang, Y. 2018,
  Phys. Rev., D97, 044017, \dodoi{10.1103/PhysRevD.97.044017}

\bibitem[{{Melatos} \& {Payne}(2005)}]{Melatos05}
{Melatos}, A., \& {Payne}, D.~J.~B. 2005, \apj, 623, 1044,
  \dodoi{10.1086/428600}

\bibitem[{Messenger {et~al.}(2015)}]{ScoX1MDC}
Messenger, C., {et~al.} 2015, Phys.\ Rev.\ D., 92, 023006,
  \dodoi{10.1103/PhysRevD.92.023006}

\bibitem[{Mukherjee {et~al.}(2018)Mukherjee, Messenger, \&
  Riles}]{Mukherjee:2017qme}
Mukherjee, A., Messenger, C., \& Riles, K. 2018, Phys. Rev. D, 97, 043016,
  \dodoi{10.1103/PhysRevD.97.043016}

\bibitem[{Nieder {et~al.}(2019)}]{Nieder:2019cyc}
Nieder, L., {et~al.} 2019, \dodoi{10.3847/1538-4357/ab357e}

\bibitem[{Nieder {et~al.}(2020)}]{Nieder:2020yqy}
---. 2020.
\newblock \doarXiv{2009.01513}

\bibitem[{\"Ozel \& Freire(2016)}]{Ozel:2016oaf}
\"Ozel, F., \& Freire, P. 2016, Ann. Rev. Astron. Astrophys., 54, 401,
  \dodoi{10.1146/annurev-astro-081915-023322}

\bibitem[{{Papaloizou} \& {Pringle}(1978)}]{Papaloizou:1978}
{Papaloizou}, J., \& {Pringle}, J.~E. 1978, MNRAS, 184, 501,
  \dodoi{10.1093/mnras/184.3.501}

\bibitem[{{Parfrey} {et~al.}(2016){Parfrey}, {Spitkovsky}, \&
  {Beloborodov}}]{Parfrey16}
{Parfrey}, K., {Spitkovsky}, A., \& {Beloborodov}, A.~M. 2016, \apj, 822, 33,
  \dodoi{10.3847/0004-637X/822/1/33}

\bibitem[{Patruno {et~al.}(2017)Patruno, Haskell, \&
  Andersson}]{Patruno:2017oum}
Patruno, A., Haskell, B., \& Andersson, N. 2017, Astrophys. J., 850, 106,
  \dodoi{10.3847/1538-4357/aa927a}

\bibitem[{{Patruno} \& {Watts}(2012)}]{Patruno12}
{Patruno}, A., \& {Watts}, A.~L. 2012, arXiv e-prints, arXiv:1206.2727.
\newblock \doarXiv{1206.2727}

\bibitem[{{Ray} {et~al.}(2019){Ray}, {Arzoumanian}, {Ballantyne}, {Bozzo},
  {Brandt}, {Brenneman}, {Chakrabarty}, {Christophersen}, {DeRosa}, {Feroci},
  {Gendreau}, {Goldstein}, {Hartmann}, {Hernanz}, {Jenke}, {Kara}, {Maccarone},
  {McDonald}, {Nowak}, {Phlips}, {Remillard}, {Stevens}, {Tomsick}, {Watts},
  {Wilson-Hodge}, {Wood}, {Zane}, {Ajello}, {Alston}, {Altamirano}, {Antoniou},
  {Arur}, {Ashton}, {Auchettl}, {Ayres}, {Bachetti}, {Balokovic}, {Baring},
  {Baykal}, {Begelman}, {Bhat}, {Bogdanov}, {Briggs}, {Bulbul}, {Bult},
  {Burns}, {Cackett}, {Campana}, {Caspi}, {Cavecchi}, {Chenevez}, {Cherry},
  {Corbet}, {Corcoran}, {Corsi}, {Degenaar}, {Drake}, {Eikenberry}, {Enoto},
  {Fragile}, {Fuerst}, {Gandhi}, {Garcia}, {Goldstein}, {Gonzalez},
  {Grefenstette}, {Grinberg}, {Grossan}, {Guillot}, {Guver}, {Haggard},
  {Heinke}, {Heinz}, {Hemphill}, {Homan}, {Hui}, {Huppenkothen}, {Ingram},
  {Irwin}, {Jaisawal}, {Jaodand}, {Kalemci}, {Kaplan}, {Keek}, {Kennea},
  {Kerr}, {van der Klis}, {Kocevski}, {Koss}, {Kowalski}, {Lai}, {Lamb},
  {Laycock}, {Lazio}, {Lazzati}, {Longcope}, {Loewenstein}, {Maitra}, {Majid},
  {Maksym}, {Malacaria}, {Margutti}, {Martindale}, {McHardy}, {Meyer},
  {Middleton}, {Miller}, {Miller}, {Motta}, {Neilsen}, {Nelson}, {Noble},
  {O'Brien}, {Osborne}, {Osten}, {Ozel}, {Palliyaguru}, {Pasham}, {Patruno},
  {Pelassa}, {Petropoulou}, {Pilia}, {Pohl}, {Pooley}, {Prescod-Weinstein},
  {Psaltis}, {Raaijmakers}, {Reynolds}, {Riley}, {Salvesen}, {Santangelo},
  {Scaringi}, {Schanne}, {Schnittman}, {Smith}, {Smith}, {Snios}, {Steiner},
  {Steiner}, {Stella}, {Strohmayer}, {Sun}, {Tauris}, {Taylor}, {Tohuvavohu},
  {Vacchi}, {Vasilopoulos}, {Veledina}, {Walsh}, {Weinberg}, {Wilkins},
  {Willingale}, {Wilms}, {Winter}, {Wolff}, {in 't Zand}, {Zezas}, {Zhang}, \&
  {Zoghbi}}]{strobex}
{Ray}, P.~S., {Arzoumanian}, Z., {Ballantyne}, D., {et~al.} 2019, arXiv
  e-prints, arXiv:1903.03035.
\newblock \doarXiv{1903.03035}

\bibitem[{{Singh} {et~al.}(2020){Singh}, {Haskell}, {Mukherjee}, \&
  {Bulik}}]{Singh20}
{Singh}, N., {Haskell}, B., {Mukherjee}, D., \& {Bulik}, T. 2020, \mnras, 493,
  3866, \dodoi{10.1093/mnras/staa442}

\bibitem[{{Suvorov} {et~al.}(2016){Suvorov}, {Mastrano}, \&
  {Geppert}}]{Suvorov16}
{Suvorov}, A.~G., {Mastrano}, A., \& {Geppert}, U. 2016, \mnras, 459, 3407,
  \dodoi{10.1093/mnras/stw909}

\bibitem[{{Suvorova} {et~al.}(2016{\natexlab{a}}){Suvorova}, {Sun}, {Melatos},
  {Moran}, \& {Evans}}]{Viterbi2016}
{Suvorova}, S., {Sun}, L., {Melatos}, A., {Moran}, W., \& {Evans}, R.~J.
  2016{\natexlab{a}}, \prd, 93, 123009, \dodoi{10.1103/PhysRevD.93.123009}

\bibitem[{{Suvorova} {et~al.}(2016{\natexlab{b}}){Suvorova}, {Sun}, {Melatos},
  {Moran}, \& {Evans}}]{ViterbiSideband}
---. 2016{\natexlab{b}}, \prd, 93, 123009, \dodoi{10.1103/PhysRevD.93.123009}

\bibitem[{{Ushomirsky} {et~al.}(2000){Ushomirsky}, {Cutler}, \&
  {Bildsten}}]{2000MNRAS.319..902U}
{Ushomirsky}, G., {Cutler}, C., \& {Bildsten}, L. 2000, Mon.\ Not.\ R.\
  Astron.\ Soc., 319, 902, \dodoi{10.1046/j.1365-8711.2000.03938.x}

\bibitem[{Vallisneri {et~al.}(2015)Vallisneri, Kanner, Williams, Weinstein, \&
  Stephens}]{Vallisneri:2014vxa}
Vallisneri, M., Kanner, J., Williams, R., Weinstein, A., \& Stephens, B. 2015,
  J. Phys. Conf. Ser., 610, 012021, \dodoi{10.1088/1742-6596/610/1/012021}

\bibitem[{{Vigelius} \& {Melatos}(2009)}]{Vigelius09}
{Vigelius}, M., \& {Melatos}, A. 2009, \mnras, 395, 1972,
  \dodoi{10.1111/j.1365-2966.2009.14690.x}

\bibitem[{Wagoner(1984)}]{Wagoner:1984pv}
Wagoner, R. 1984, Astrophys. J., 278, 345, \dodoi{10.1086/161798}

\bibitem[{{Wang} {et~al.}(2018){Wang}, {Steeghs}, {Galloway}, {Marsh}, \&
  {Casares}}]{ScoX1ParamRev}
{Wang}, L., {Steeghs}, D., {Galloway}, D.~K., {Marsh}, T., \& {Casares}, J.
  2018, MNRAS, 478, 5174, \dodoi{10.1093/mnras/sty1441}

\bibitem[{Watts {et~al.}(2008)Watts, Krishnan, Bildsten, \&
  Schutz}]{Watts:2008qw}
Watts, A., Krishnan, B., Bildsten, L., \& Schutz, B.~F. 2008, Mon.\ Not.\ R.\
  Astron.\ Soc., 389, 839, \dodoi{10.1111/j.1365-2966.2008.13594.x}

\bibitem[{{Watts}(2012)}]{Watts12}
{Watts}, A.~L. 2012, \araa, 50, 609,
  \dodoi{10.1146/annurev-astro-040312-132617}

\bibitem[{{Watts} {et~al.}(2019){Watts}, {Yu}, {Poutanen}, {Zhang},
  {Bhattacharyya}, {Bogdanov}, {Ji}, {Patruno}, {Riley}, {Bakala}, {Baykal},
  {Bernardini}, {Bombaci}, {Brown}, {Cavecchi}, {Chakrabarty}, {Chenevez},
  {Degenaar}, {Del Santo}, {Di Salvo}, {Doroshenko}, {Falanga}, {Ferdman},
  {Feroci}, {Gambino}, {Ge}, {Greif}, {Guillot}, {Gungor}, {Hartmann},
  {Hebeler}, {Heger}, {Homan}, {Iaria}, {Zand}, {Kargaltsev}, {Kurkela}, {Lai},
  {Li}, {Li}, {Li}, {Linares}, {Lu}, {Mahmoodifar}, {M{\'e}ndez}, {Coleman
  Miller}, {Morsink}, {N{\"a}ttil{\"a}}, {Possenti}, {Prescod-Weinstein}, {Qu},
  {Riggio}, {Salmi}, {Sanna}, {Santangelo}, {Schatz}, {Schwenk}, {Song},
  {{\v{S}}r{\'a}mkov{\'a}}, {Stappers}, {Stiele}, {Strohmayer}, {Tews},
  {Tolos}, {T{\"o}r{\"o}k}, {Tsang}, {Urbanec}, {Vacchi}, {Xu}, {Xu}, {Zane},
  {Zhang}, {Zhang}, {Zhang}, {Zheng}, \& {Zhou}}]{extp}
{Watts}, A.~L., {Yu}, W., {Poutanen}, J., {et~al.} 2019, Science China Physics,
  Mechanics, and Astronomy, 62, 29503, \dodoi{10.1007/s11433-017-9188-4}

\bibitem[{{Whelan} {et~al.}(2015){Whelan}, {Sundaresan}, {Zhang}, \&
  {Peiris}}]{LMXBCrossCorr}
{Whelan}, J.~T., {Sundaresan}, S., {Zhang}, Y., \& {Peiris}, P. 2015, Phys.\
  Rev.\ D., 91, 102005, \dodoi{10.1103/PhysRevD.91.102005}

\bibitem[{{Woan} {et~al.}(2018){Woan}, {Pitkin}, {Haskell}, {Jones}, \&
  {Lasky}}]{Woan18}
{Woan}, G., {Pitkin}, M.~D., {Haskell}, B., {Jones}, D.~I., \& {Lasky}, P.~D.
  2018, \apjl, 863, L40, \dodoi{10.3847/2041-8213/aad86a}

\bibitem[{Zhang {et~al.}(2020)Zhang, Papa, \& Krishnan}]{O2CrossCorr-AEI}
Zhang, Y., Papa, M.~A., \& Krishnan, B. 2020, {Supplemental materials to the
  paper {\it{Search for Continuous Gravitational Waves from Scorpius X-1 in
  LIGO O2 Data}}}, \url{www.aei.mpg.de/continuouswaves/CrossCorr-O2-20-180}

\end{thebibliography}
\bibliographystyle{aasjournal}

\end{document}